**A Case Study on Job Scheduling Policy for Workload Characterization and Power Efficiency**

A. A. CHANDIO++, Z. YU, F. S. SYED**, I. A. KOREJO*

Shenzhen Institutes of Advanced Technology, Chinese Academy of Sciences, China.

**Abstract:** With the increasing popularity of cloud computing, datacenters are becoming more important than ever before. A typical datacenter typically consists of a large number of homogeneous or heterogeneous servers connected by networks. Unfortunately, these servers and network equipment are often under-utilized and power hungry. To improve the utilization of hardware resources and make them power efficiency in datacenters, workload characterization and analysis is at the foundation. In this paper, we characterize and analyze the job arriving rate, arriving time, job length, power consumption, and temperature dissipation in a real world datacenter by using statistical methods. From the characterization, we find unique features in the workload can be used to optimize the resource utilization and power consumption of datacenters.

## 1. INTRODUCTION

Recently, there is an increasing interest on cloud computing and it has become a clear trend of information technology over the world. Cloud computing providers such as Amazon (Amazon online 2008), Google, and IBM has already built large scale cloud platforms and started to provide a wide range of services such as computing resources renting and online documents to end users including companies and individual persons. The most important part of cloud computing is datacenter because it is the host of all cloud services. A typical datacenter consists of tens to hundreds of thousands of homogeneous or heterogeneous servers connected by networks. Unfortunately, these servers and network equipments are often under-utilized and power hungry. Hence, most of computing resources are wasted while continuously consuming power.

Cloud Computing support the large number of users to different organizations on the same time. Other hand, Resource Provider (RP) provides the non-infinite resources. For this reason RPs are making attention on resources management for own satisfaction and the demand of organization. There are two major concentrated points rising in RPs mind; to control the overall energy consumption on Datacenter and to response the deployed organization in satisfied time of period. The energy consumption is the largest operational cost for these environments. In recent years, RPs is growing day by day such as IBM, Google and Microsoft has deployed the Datacenters for scientific research as well as hosting of Internet applications. Typically the size of these Datacenters is thousands of thousands servers and various components. And all these components consumes different amount of electricity for their operations as well as for cooling system. The breakdown of peak power consumed by major components of single server with 2 x86 CPU, an IDE disk drive is shown in Table I. taken from (Fan *et al.* 2007) and the CPU is one largest power consume component than other.

**Table-1 Component Peak Power Breakdown For A Typical Server**

| Component | Peak | Count | Total |
|---|---|---|---|
| CPU | 40 W | 2 | 80 W |
| Memory | 9 W | 4 | 36 W |
| Disk | 12 W | 1 | 12 W |
| PCI slots | 25 W | 2 | 50 W |
| Motherboard | 25 W | 1 | 25 W |
| Fan | 10 W | 1 | 10 W |
| **System Total** | | | **213 W** |

In analysis report (Brown 2008) for energy use and cost in Datacenter, this obtained by US Environment Protection Agency (EPA) in 2007. The Datacenter was consumed 1.5% of total sales of electricity in country for the year 2006, that was 61 billion kilowatt-hour (kWh). Also it was estimated as double, i.e., more than 100 billion kWh for 2011. According (Koomey 2008), worldwide electricity in Datacenter is aggregated as double from 2000 to 2005. About 80% of this growth was attributable to growth in electricity used by servers. About 10% associated with Datacenter communication and about the same percentage only for storage equipment. It could be decrease the overall percentage of power consumptions in Datacenter via implementations of strong techniques and more research concentrate on particular this

++Corresponding Author: A. A. Chandio, aftabac@siat.ac.cn , Cell No. +86 13244762252
*Institute of Mathematics & Computer Sciences, University of Sindh, Jamshoro, Pakistan.
**Mehran University of Engineering and Technology.

direction. In 2011 another analysis report of the same author described the rapid rates of growth in Datacenter electricity use. That it prevailed from 2000 to 2005 slowed significantly from 2005 to 2010 instead of doubling because of further improvement in virtualization and other factors (Koomey 2011). To apply the Dynamic Voltage and Frequency Scaling (DVFS), to improve the server, storage and cooling efficiency, Dynamic Power Management (DPM) or device reduction, Multi-core processors designs and virtualization are the major techniques to reduce the power consumption (Brown 2008 and *Junaid. et al.,* 2012). Author in (Kommey 2011) summarized via analysis a total electricity used by Datacenters in 2010 of about 1.3% of all electricity used for the world, and 2% of all electricity used for the US.

Adjacent to concentrate on the energy consumption, RPs also have to take strong efforts for response the organization and fulfill the different Quality of Service (QoS) constraints. Then resources might be highly under-utilized and this situation could generate high temperature on Datacenter which will be a cause of hardware unreliability and performance deficiency. Such growth of Datacenters in cloud environment leads for in-depth studies of energy efficiency as well load efficiency. However it has big tie between both goals. There are major techniques discussed in this paper, to minimize the temperature on Datacenter, including DVFS, DPM and job scheduler; our focus only on job scheduling technique, because in programming model via an appropriate scheduling policy can be achieve the ability to utilize the resources as well as to reduce energy consumption on processor.

It is necessary for these assumptions to optimize the system for finding the overall load and power consumption on Datacenter. Once RPs knows the fine-grain information about load and its temperature contribution then through scheduler, it could be achieved optimal results for resources utilization under the energy efficiency. Workload characterization gives us motivation to interpret the difference between job computation times. Based on which the schedulers are able to schedule the frequency. We use a real Datacenter workload for characterization by using statistical analysis methods and collect insight information of the system behavior. Then we will simulate the real Datacenter workload by implementation of well-known scheduler, which will be in our future work. We characterize the workloads according two phases; before the scheduling and after the scheduling. The workload characteristics before the scheduling process are the independent and the characteristics after scheduling depend on the scheduler policy, which could be found with different characteristics in different scheduling policy. This paper only analyzes the default scheduling policy. We find very important findings including job's size and node's temperature.

The rest of this paper is arranged as follows. Section 2 abides energy efficiency techniques in Datacenter, Section 0 address the characterization of Datacenter workloads, in Section 4 we analyze the job profile, node profile and temperature profile which are affected by a default scheduler and Section 0 shows the conclusion.

## 2.   TECHNIQUES FOR ENERGY EFFICIENCY

As primary focus of Datacenter is a better performance and high throughput, the power efficiency is also important to consider in Datacenter constructions. Many techniques can be applied for energy efficiency in Datacenter without maximum overheads. Such as, to apply the Dynamic Voltage and Frequency Scaling (DVFS), to improve the server, storage and cooling efficiency, Dynamic Power Management (DPM) or device reduction, Multi-core processors designs and virtualization are the major techniques to reduce the energy consumption (Brown 2008 and Junaid. et al 2012). In DPM (Benini *et al.* 2000), the devices could be powered on/off dynamically according to currant load. The DVFS technique (Weiser *et al.* 1994) dynamically changes the voltage and frequency of devices. Therefore, by voltage and frequency reduction on CPU or switch boards, the power consumption can be reduced. In virtualization, one physical server is shared via multiple virtual machines (VM), and these VM dynamically provisioned the resources such as CPU and memory according demands. All these techniques could be fixed in job scheduling strategy for better performance. Job scheduling strategy is best technique for energy efficiency and resources utilization.

### *A. Job Scheduling Strategy*

As large changing in the processors development is increasing, such as, uniprocessor to multiprocessor and single core to many core and then multicore. Therefore, programming models must be more human-centric and engage the full range of issues associated with developing a parallel application on hardware. This development in hardware need to simplify the programming model to efficiently program to make parallel computing productive, scalable and energy efficient (Asanovic *et al.* 2006). The job scheduling policy is a major component of resources managements in large-scale parallel computing environment. The general concept of job scheduling policy is to select the processing resources for job submitted and to find the perfect job for idle processing resources. These jobs may be shuffled in any order or time for giving the priority to jobs. Such as largest job first (LJF) scheduler shuffles the jobs in decreasing order of job's size or shortest job first (SJF) scheduler rearrange the jobs in increasing order. This operation frequently invoke either in particular time interval or at the time of new job submission. There is no doubt, some scheduling policies has some sort of limitations for example first come first serve (FCFS) does not utilize the resources and maximize the response time and SJF/LJF making long delay for non-priority jobs. To remove the limitations and to enhance the capabilities of schedulers, these could be updated via implementation of extra techniques. Such as first fit (FF) finds the job from ordered queue list, which can be fit on first idle resources; backfilling technique (Mu'alem and Feitelson 2001) makes the resources reservation for job and backfill the job if second job which is small longer than first could be execute in the time portion which could not violated the time reserved for first job. FF technique is suitable in SJF, LJF as well in first come first serve (FCFS) classical scheduling policies where the job served on first come basis and backfilling technique was developed for IBM SP1 parallel supercomputer on FCFS scheduler basis (Mu'alem and Feitelson 2001). Advance knowledge of workload and overall job's characteristics for system, can make easy to select or update the scheduler.

In programming model, the scheduler does not only provide the ability to utilize the resources but also large space for programming techniques to reduce energy consumption on processor. According to (Vanderster *et al.* 2007) different tasks generate the different amount of heat and create distinct task-temperature profile. Long-time execution of job generates the more temperature as to compare short-time execution of job. To maintain the load on processor is a major key to control the power consumptions because high load makes processor overheated and forced to slow down execution. Then all vital measures will be degraded, such as throughput and utilization will be reduced, response time will increase, jobs are more likely to miss deadlines, etc. Thus processor overheating will negatively affect the overall system performance (Yang *et al.* 2008). A scheduling policy could estimate the job's temperature from execution time is predefined by user and schedule the job according its priority. Such priority for reduce the temperature likewise schedule the hot job before the cool one and results will be in a lower final temperature (Yang et al. 2008). The performance and temperature of processor depend on the order of job execution in scheduler according to job characteristics such as heat contribution (Chrobak *et al.* 2008).

### 3. CHARACTERIZATION OF DATACENTER WORKLOAD

Resources management system in Datacenter stores the information about past usage of the system. This information is found in a log format, which called workload. These logs provide information about every job execution in the system. The performance of system is evaluated by characteristics of hardware and software components as well as of the load it has to process (Calzarossa and Serazzi 1993, and Feitelson 2002). Therefore, workload characterization is playing an important role in system design especially for job schedulers, it allows scheduler to understand the overall behave between user and system. In which arrival rate job comes from user, what different size of jobs, what length of job, all the jobs related questions divert the administrator mind. Then some system related questions rise such as how to manage system under load, how to utilize the resources, how to make user satisfaction (i.e., QoS constraints) and how to minimize the total cost of ownership (TCO). These questions lead to administrator for redesign the system via appropriate techniques.

We characterize the real Datacenter workload in this paper. In order to understand the workload of Datacenter, it is required to have best knowledge about itself Datacenter. Datacenter is collection of multiple servers constructed on computational resources and depends upon the different nature, such as the resources could be categorized in two different categories, Homogeneous and Heterogeneous. Homogeneous systems perform the execution of task in similar capacity due to its all resources similarity such as the CPU speed, memory capacity, cache size, etc. Unlikely in homogeneous environment, the heterogeneous parallel system is collection of resources with different size and capacity, and allocates the tasks on various processors with different speed. With the advantage of Datacenter in cloud computing, it responds to more number of users with pay-and-use and pay-as-you-go methods known utility computing, it runs several jobs on the same time. These jobs may belong to complex problems such as computation intensive, taking large time for execution, etc.

### *B. Log's information*

We use the real world data from a production Datacenter. The workload is collected during 30 days' time period from 20 Feb. 2009 to 22 Mar. 2009. It contains 22385 jobs ran out on more than 1000 nodes. Each node in Datacenter has two CPUs and Datacenter contains homogenous set of resources. In log files, the data comes from three different tables. The information in these tables came before and after the execution of default scheduler. *Job-data* table stored job information such as what time job was requested to scheduler, what time job was executed on CPU, what number of resources was demanded by job and what time job was finished the execution. *Job-nodes* table stored the information of jobs and resources that on which resources the job was executed. *Temp-data* table stored each node's temperature and it was hourly captured.

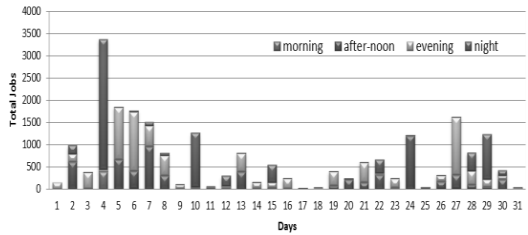

**Fig. 1.** Distribution of Job arriving before the schedule,

### C. Job information

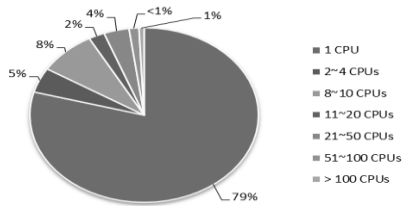

**Fig. 2.** Job's breakdown according number of CPUs.

Job is a program, which is submitted by user on particular time. Each job has multiple tasks and each task execute on separate CPU for time period. In this section we characterize the job to find out hidden information such as job arrival rate on specific time and job size. Fig.1 shows the hourly job arrival rate for 30-day. Few hours clear shows the high load arriving.

In (**Fig. 2**) we distribute the job's arriving rate per day. In this graph we create categorical variables for each day. The 24 hours in each day is divided in four categorical variables including morning (6 AM to 11 AM), afternoon (12 PM to 5 PM), evening (6 PM to 11 PM) and night (12 AM to 5 AM). Here we found the large number of jobs arrived in fourth day especially in night time. In all 30-day maximum jobs arrived in evening and night time such as 36.8 % and 32.3 % respectively. In both findings, the data do not have any trend in particular time cycle. Therefore the arrival rate of workloads is totally variable.

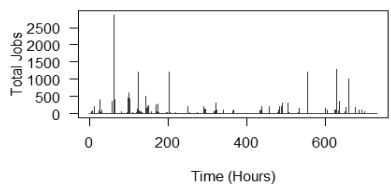

**Fig. 3.** Job arrival rate on Datacenter per

In the next observation, we analyze the job distribution via its size. As job has rectangle shape (Srinivasan *et al.* 2002), we summarized the job size in two perspectives; job width and job length, shows the size in account of resources and execution time, respectively. (**Fig. 3**), defines the breakdown of Jobs according CPU consumption. We found that all the jobs demand an even number of CPUs or single CPU for execution. The maximum percentage of jobs demands the single CPU for execution. (**Fig. 3**) shows the pie chart for job distribution in job width point of view.

Job length is second perspective for distribution of job size, where we use the job execution time (running time) which is, how long time job remains on processor for execution. We need to find whether jobs are CPU computation intensive. In Table II., there are about half percent of total jobs within an hour long execution and half percent of total jobs with long execution more than an hour. The maximum number of within an hour longer jobs only presents in the category of less than 20 minutes longer executed jobs and the maximum number of more than an hour longer jobs presents in the range of 12 hours. In Table III we distribute the jobs size for correlation between width and length. The jobs with 1 CPU size are more dominant than others in all categories of job length shown in the table, and between 11 to 20 minutes and more than 9 hours longer jobs are more dominant than others categories of job length such as 27 % and 20 %, respectively.

**Table 2. Breakdown Distribution of Job Length**

| Job length | No. of jobs | % of jobs |
|---|---|---|
| 1 hour | 10428 | 46.58 |
| >1 hour | 11957 | 53.42 |
| Total | 22385 | 100 |

**Table 3. Percentage Breakdown For Correlation Between Jobs Width And Length**

| Job size | 1 | 2~24 | > 32 | Total |
|---|---|---|---|---|
| < 1 Min. | 4.83 | 0.58 | 0.13 | 5.55 |
| 2~10 | 7.60 | 0.90 | 0.38 | 8.88 |
| 11~20 | 20.72 | 3.40 | 2.97 | **27.09** |
| 21~60 | 3.73 | 1.07 | 0.27 | 5.07 |
| 2~4 hours | 13.23 | 0.67 | 0.68 | **14.57** |
| 5~8 hours | 18.43 | 0.15 | 0.16 | **18.74** |
| > 9 hours | 10.73 | 8.27 | 1.10 | **20.11** |
| **Total** | **79.27** | 15.03 | 5.70 | 100 |

### 4. ANALYZING THE SCHEDULER RESULTS

In this section, we analyze to default scheduler policy in order to find out the problems and system behave. The scheduler assigns the jobs to idle machines in order to achieve the best resources utilization and load balancing. We use the statistical methods for job distribution on nodes and the temperature of nodes. When scheduler schedules the jobs, it would effect on node profile and temperature profile. We analyze the scheduler result in different way; node-profile and temperature-profile.

### D. Node profile

Frist observation to know the load balance on machines, we analyze a complete distribution of jobs on each node in Datacenter. This distribution is created after execution of default scheduler. **Fig. 4 (a)** shows the job distribution on each node scheduled by default policy. The minimum number of jobs 3 and maximum 429 numbers of jobs ran out on a node. The first and third quartile is 89 and 148, respectively. There are about 20 nodes shown as outliers in **Fig 4 (b)** box plot, starting from 250, and range is only going to maximum direction, which may be a cause of large number of jobs executed on only 20 nodes. Consequently, from this box plot, we can observe the minimum and maximum range as 3 and 250, and a mean is only 121 numbers of jobs. The nodes with large number of jobs may be executed

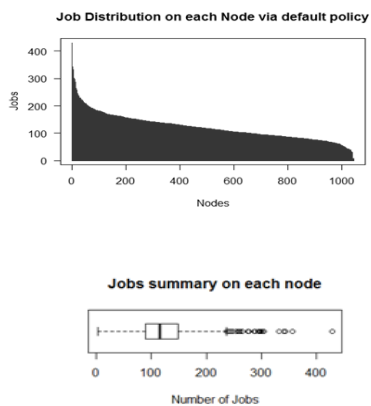

**Fig. 4.** Job distribution on each node via default scheduler policy (a) job count on each node (b) summary of each nod.

shorter jobs and the nodes with small number of jobs may be executed longer jobs.

### E. Temperature profile

After workload profile, our next analysis in this section is overlooking on temperature profile of each node in Datacenter. Temperature profile is captured after every hour in 30-day. We found, the maximum and minimum temperature on nodes 131 and 71 $F^0$, respectively. We have also categorized the temperature profile in some categories, shown in bar chart in Figure 5, where we can see the correlation between nodes in particular category. Maximum number of intervals has been found the temperature in the range of 95 to 104 $F^0$.

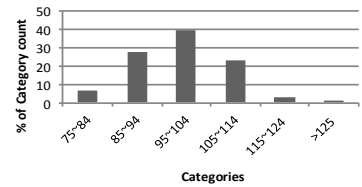

**Fig. 5.** Temperature categorized.

In further analysis we found that the idle state of a node can have at least temperature. In default policy, if a node is in idle state also generate the least temperature, it is shown in below Figure 6 (a) for load on a node and Figure 6 (b) for temperature of same node. This figure shows clearly that the even zero loads on a node create temperature in range 75 to 84 $F^0$.

## 5. CONCLUSION

In this paper we characterized the Datacenter workload for optimization as we can find the overall Datacenter load efficiency and energy efficiency (DCLE/EE). We use the statistic for workload characterization. In this paper we find the job's and node's behavior before the scheduler as well as after the scheduler. From the characterization, we find unique features in the workload: i) maximum percentage of jobs demands the single CPU for execution; ii) remaining jobs demands the even number of CPU for execution; iii) half percent of total jobs executes as long within an hour; iv) half executes as long more than an hour. And the results after default scheduler; i) 40 % of total intervals (hours) of nodes are found in range of 95 to 104 Fahrenheit degree of temperature in an idle state of a node.

We confirm that in our future work, all above observations would be helpful to redesign the scheduler policy in order to achieve both energy and load efficiency in Datacenter.

### ACKNOWLEDGEMENT

This is an extended version of the paper "A Characterization and Analysis Data Center Workloads" published as conference proceedings of International Conference on Computers and Emerging Technologies (ICCET 2013) held on March 5-7, 2013, organized by Department of Computer Science, Shah Abdul Latif University Khairpur Sindh Pakistan.